\documentclass[aps,prd,superscriptaddress,twocolumn,secnumarabic,nofootinbib]{revtex4-1}
\usepackage{amsmath}
\usepackage{amsfonts}

\pdfoutput=1

\usepackage{bm}
\usepackage{times}
\usepackage{braket}
\usepackage{color,graphicx}
\usepackage[utf8]{inputenc}
\usepackage[T1]{fontenc}
\newcommand{\cb}{{\cal{B}}}
\newcommand{\bea}{\begin{eqnarray}}
\newcommand{\eea}{\end{eqnarray}}

\newcommand{\E}[1]{\times 10^{#1}}

\usepackage[normalem]{ulem}

\definecolor{nicered}{rgb}{0.7,0.1,0.1}
\definecolor{nicegreen}{rgb}{0.1,0.5,0.1}

\usepackage{hyperref}
\hypersetup{colorlinks,citecolor=nicegreen,linkcolor=nicered}
\hypersetup{colorlinks=true}

\begin{document}
\title{On a model with two scalar leptoquarks - $R_2$ and $S_3$}

\author{Damir Be\v cirevi\' c} \email{damir.becirevic@ijclab.in2p3.fr}
\affiliation{IJCLab, Pôle Théorie (Bât.~210), CNRS/IN2P3 et Université Paris-Saclay,
91405 Orsay, France}

\author{Ilja Dor\v sner} \email{dorsner@fesb.hr}
\affiliation{University of Split, Faculty of Electrical Engineering, Mechanical Engineering and Naval Architecture in Split (FESB), Ru\dj era Bo\v skovi\' ca 32, 21 000 Split, Croatia}

\author{Svjetlana Fajfer} \email{svjetlana.fajfer@ijs.si}
\affiliation{Department of Physics, University of Ljubljana, Jadranska 19, 1000 Ljubljana, Slovenia}
\affiliation{Jo\v zef Stefan Institute, Jamova 39, P.\ O.\ Box 3000, 1001
  Ljubljana, Slovenia}

\author{Darius A.~Faroughy}
\email{faroughy@physik.uzh.ch}
\affiliation{Physik-Institut, Universit\"at Z\"urich, CH-8057 Z\"urich, Switzerland}
\author{Florentin Jaffredo}
\email{florentin.jaffredo@ijclab.in2p3.fr}
\affiliation{IJCLab, Pôle Théorie (Bât.~210), CNRS/IN2P3 et Université Paris-Saclay,
91405 Orsay, France}
\
\author{Nejc Ko\v snik} \email{nejc.kosnik@ijs.si}
\affiliation{Department of Physics, University of Ljubljana, Jadranska 19, 1000 Ljubljana, Slovenia}
\affiliation{Jo\v zef Stefan Institute, Jamova 39, P.\ O.\ Box 3000, 1001
  Ljubljana, Slovenia}
  
\author{Olcyr Sumensari} \email{olcyr.sumensari@ijclab.in2p3.fr}
\affiliation{IJCLab, Pôle Théorie (Bât.~210), CNRS/IN2P3 et Université Paris-Saclay,
91405 Orsay, France}

\begin{abstract}
We discuss a model that can accommodate the $B$-physics anomalies, based on combining two scalar leptoquarks, $R_2$ and $S_3$, of mass $\mathcal{O}(1\,\mathrm{TeV})$, and that we proposed in our previous paper. We update the analysis of its parameter space and show that a model remains viable and consistent with a number of low energy and high energy flavor physics constraints. Since the model predicts a non-zero New Physics phase, we discuss the possibility to test its contribution to the neutron electric dipole moment and to the angular distributions of the exclusive $b\to c\tau \bar \nu$ decays. We find that the model can provide a significant enhancement to $\mathcal{B}(B\to K^{(\ast)}\nu\nu)$ and provides both the upper and lower bounds to $\mathcal{B}(B\to K^{(\ast)}\mu\tau)$.
\end{abstract}
\maketitle

\section{Introduction}
\label{sec:intro}
It has been established that the only viable scenario involving a single $\mathcal{O}(1~\mathrm{TeV})$ mediator 
that can accommodate the so called $B$-anomalies, while remaining consistent with a wealth of measured low energy flavor observables, 
as well as with the bounds arising from the direct searches at the LHC and those deduced from the high-$p_T$ tails of $pp \to \ell \ell$,  
 is a scenario with a singlet vector leptoquark (LQ) state, often referred to as $U_1 = (\bm{3},\bm{1},2/3)$, where the quantum numbers correspond to the Standard Model (SM) gauge group~\cite{Angelescu:2018tyl}. While the vector LQ is an appealing solution, it creates problems when building a particular model because the resulting effective theory is not renormalizable unless a particular ultra-violet (UV) completion to the theory is specified~\cite{DiLuzio:2017vat}. This, in turn, necessitates introducing more states, more parameters, and in order to make a model more predictive a number of assumptions needs to be made. An alternative to that scenario is to combine two scalar LQ's, such as $S_1= (\overline{\bm{3}},\bm{1},1/3)$ with $S_3= (\overline{\bm{3}},\bm{3},1/3)$~\cite{S13}, or $R_2= (\bm{3},\bm{2},7/6)$ with $S_3$. The advantage of the two scalar LQ scenarios is that they remain renormalizable, the loop processes can be easily computed without necessity of introducing an UV cutoff by hand, so that, once measured, such processes can be used as constraints. In Ref.~\cite{Becirevic:2018afm} a model in which the $R_2$ leptoquark is combined with $S_3$ both with mass $\mathcal{O}(1~\mathrm{TeV})$ has been proposed. In order to make it minimalistic we chose the structure of Yukawa couplings~\footnote{Yukawa couplings here are couplings between LQ's and particular quark and lepton flavors.} such that the matrices of left-handed couplings to $R_2$ and to $S_3$ are related via $Y_L^{(S_3)} = -Y_L^{(R_2)}$, a pattern that can provide a plausible embedding of the resulting effective theory in a $SU(5)$ unification scenario.

In this paper we update the analysis presented in Ref.~\cite{Becirevic:2018afm} to show that the proposed scenario is still viable and consistent with the current experimental data. Furthermore, we discuss several new observables, including those relevant to the angular distributions of $B\to D^\ast (\to D\pi) \tau\bar \nu$ and $\Lambda_b\to \Lambda_c (\to \Lambda\pi) \tau\bar \nu$, the measurement of which can help distinguishing this particular model from the other ones proposed in the literature. Another novelty is the analysis of the high-$p_T$ tails both of the mono-$\tau$ and di-$\tau$ events for various leptoquark masses, which has not been discussed in our previous paper~\cite{Becirevic:2018afm}.

\section{Model \label{sec:1}}

As mentioned above, we combine  $R_2$ with $S_3$ LQs in order to accommodate both kinds of $B$-anomalies. More specifically, the observation that the (partial) branching fractions of the exclusive $b\to s\mu\mu$ processes are smaller than predicted in the SM can be described by couplings to a  $S_3$ LQ, whereas those showing the excess of events based on the $b\to c\tau\bar\nu$ transition can be described by couplings to $R_2$. 

To be more specific, the interaction Lagrangian between the LQ's and the SM fermions in this model reads:
\bea
\label{eq:one}
\mathcal{L}  &\supset & \left(Y_{R}^{(R_2)}\right)^{ij} \bar{Q}^\prime_i \ell^\prime_{Rj} R_2+ \left(Y_{L}^{(R_2)}\right)^{ij} \bar{u}^\prime_{Ri} \widetilde{R}_2^\dagger L^\prime_j \nonumber\\
&& + \left(Y_{L}^{(S_3)}\right)^{ij} \bar{Q}^{\prime C}_{i} i \tau_2 ( \tau_k S^k_3) L^\prime_{j} \ +\ \mathrm{h.c.},
\eea
where $Y_L^{(R_2,S_3)}$ and $Y_R^{(R_2)}$ are the Yukawa matrices, $\tau_k$ are the Pauli matrices, $S_3^k$ stands for a component of the $SU(2)_L$ triplet. In the above expression we use the notation with $\widetilde{R_2} = i \tau_2 R_2^\ast$. In the mass eigenstate basis the above Lagrangian becomes:
\begin{align}
\label{eq:two}
\begin{split}
\mathcal{L} \supset 
&+(V Y_R^{(R_2)} E_R^\dagger)^{ij} \bar{u}_{Li}\ell_{Rj}R_2^{\frac{5}{3}} + (Y_R^{(R_2)} E_R^\dagger)^{ij} \bar{d}_{Li}\ell_{Rj} R_2^{\frac{2}{3}}\\
&+(U_R Y_L^{(R_2)} U)^{ij} \bar{u}_{Ri} \nu_{Lj} R_2^{\frac{2}{3}}- (U_R Y_L^{(R_2)})^{ij} \bar{u}_{Ri}\ell_{Lj} R_2^{\frac{5}{3}}\\
&-(Y_L^{(S_3)} U)^{ij} \bar{d}^C_{Li} \nu_{Lj} S_3^{\frac{1}{3}} +\sqrt{2}(V^* Y_L^{(S_3)} U)^{ij} \bar{u}^C_{Li} \nu_{Lj} S_3^{-\frac{2}{3}}\\
&- \sqrt{2}  (Y_L^{(S_3)})^{ij} \bar{d}^C_{Li} \ell_{Lj} S_3^{\frac{4}{3}} -(V^* Y_L^{(S_3)})^{ij} \bar{u}^C_{Li} \ell_{Lj} S_3^{\frac{1}{3}} \\ 
&+\ \mathrm{h.c.},
\end{split} 
\end{align}
where the superindices in $R_2$ and $S_3$ now refer to the electric charge. In what follows we will assume the components of the $R_2$ doublet 
and those of the $S_3$ triplet to be mass degenerate, respectively.  In our notation, the mass and flavor eigenstates are related via 
$u_{L,R}=U_{L,R} u^\prime_{L,R}$, $d_{L,R}=D_{L,R} d^\prime_{L,R}$,
$\ell_{L,R}=E_{L,R} \ell^\prime_{L,R}$, $\nu_{L}=N_L \nu^\prime_{L}$, where
$U_{L,R}$,
$D_{L,R}$, $E_{L,R}$, and $N_L$ are unitary matrices. Therefore, $V= U_L D_L^\dagger \equiv U_L$ and
$U=E_L N_L^\dagger \equiv N_L^\dagger$ are the CKM and the PMNS
matrices, respectively.

Concerning the Yukawa matrices we assume their structure to be minimalistic and the non-zero values are:
\begin{equation}
\label{eq:yL-yR}
Y_R^{(R_2)} E_R^\dagger = \begin{pmatrix}
0 & 0 & 0\\ 
0 & 0 & 0\\ 
0 & 0 & y_R^{b\tau}
\end{pmatrix},~ 
U_R Y_L^{(R_2)} = \begin{pmatrix}
0 & 0 & 0\\ 
0 & y_L^{c\mu} & y_L^{c\tau}\\ 
0 & 0 & 0
\end{pmatrix}\,,
\end{equation}
where, as mentioned above, we take $Y_L^{(S_3)}=-Y_L^{(R_2)}$, namely,
\begin{equation}
\label{eq:yL-S3}
Y_L^{(S_3)} =-  \begin{pmatrix}
1 & 0 & 0\\ 
0 & \cos\theta & \sin\theta\\ 
0 & -\sin\theta & \cos\theta
\end{pmatrix} \begin{pmatrix}
0 & 0 & 0\\ 
0 & y_L^{c\mu} & y_L^{c\tau}\\ 
0 & 0 & 0
\end{pmatrix}\,. 
\end{equation}
In summary, the New Physics (NP) parameters in this model are: $m_{R_2}$, $m_{S_3}$, $y_R^{b\tau}$, $y_L^{c\mu}$, $y_L^{c\tau}$, and $\theta$. All of the mentioned parameters are real except for $y_R^{b\tau}$ which we allow to be complex for the reason that will soon become clear.

\subsection{$b \to c \tau \bar \nu$}

At low energies, the above model, when applied to describing the $b\to c\tau \bar \nu$ processes, reduces to the effective theory 
\begin{equation}
\begin{split}
{\cal H}^{b \to c \tau \bar \nu}_{\mathrm{eff}} \supset & \frac{4 \, G_F}{\sqrt{2}} V_{cb}\biggl[  g_{S_L}(\mu)\, (\bar{c}_R  b_L)(\bar{\tau}_R \nu_{L}) \\
&+ g_T(\mu)\, (\bar{c}_R  \sigma_{\mu \nu} b_L) (\bar{\tau}_{R} \sigma^{\mu \nu}\nu_L)
\biggr]+\ \mathrm{h.c.},
\label{eq:hamiltonian-semilep}
\end{split}
\end{equation}
which is to be added to the SM. The effective couplings that appear in ${\cal H}^{b \to c \tau \bar \nu}_{\mathrm{eff}}$  can be easily identified as
\bea\label{eq:gst}
g_{S_L}(\Lambda)= 4\, g_T(\Lambda) = \frac{ y_L^{c\tau}\  y_R^{b\tau\,\ast} }{4\sqrt{2} G_FV_{cb}\, m_{R_2}^2}\,,
\eea
at the scale $\mu = \Lambda \simeq m_{R_2}$. That relation, due to the renormalization group running, from $\Lambda \simeq 1\,\mathrm{TeV}$ down to the low energy scale $\mu = m_b$, translates to $g_{S_L}(m_b)\approx 8.1\times g_T(m_b)$~\cite{Gonzalez-Alonso:2017iyc}.~\footnote{More specifically, the relation $g_{S_L}(\Lambda)= 4\, g_T(\Lambda)$ gets modified due to renormalization group running from $\Lambda\simeq 1\,\mathrm{TeV}$ down to $\mu =m_b$. Since $g_{S_L}(m_b) =1.56\, g_{S_L}(\Lambda)$, and $g_{T}(m_b) =0.77\, g_{T}(\Lambda)$, one then obtains $g_{S_L}(m_b)\approx 8.1\times g_T(m_b)$.}
In Ref.~\cite{Becirevic:2018afm} we explicitly wrote the contribution to $b\to c\tau \bar \nu$ coming from $S_3$, which however is tiny in this scenario and will be neglected in the following discussion. 

\subsection{$b \to s  \mu \mu$}

Another type of anomalies, namely those relevant to the exclusive processes based on $b\to s\mu \mu$, are described in this framework by 
\begin{equation}
\label{eq:Heff2}
 \mathcal{H}^{b\to s \mu\mu}_{\mathrm{eff}} \supset -\dfrac{4 G_F\lambda_t}{\sqrt{2}}  \biggl[ \delta C_9(\mu)\mathcal{O}_9(\mu) + \delta C_{10}(\mu)\mathcal{O}_{10}(\mu)\biggr]+\ \mathrm{h.c.},
\end{equation}
where $\lambda_t = V_{tb}V_{ts}^\ast$, and 
\bea
&&  \mathcal{O}_{9} = \dfrac{e^2}{(4\pi)^2} \,\big{(}\bar{s}_L\gamma_\mu  b_L\big{)} \big{(}\bar{\mu}\gamma^\mu\mu \big{)}\,,\cr
&&  \mathcal{O}_{10} = \dfrac{e^2}{(4\pi)^2} \,\big{(}\bar{s}_L\gamma_\mu  b_L\big{)} \big{(}\bar{\mu}\gamma^\mu \gamma^5 \mu \big{)}\,.
\eea
After matching the amplitude obtained by using the Lagrangian~\eqref{eq:two} with the one based on the low energy effective theory~\eqref{eq:Heff2}, one finds:  
\begin{align}
\label{eq:sin2theta}
\delta C_{9}=-\delta C_{10} &= \frac{\pi v^2}{\lambda_t \alpha_\mathrm{em} } \frac{ \left( Y_L^{(S_3)}\right)^{b\mu } \left( Y_L^{(S_3)}\right)^{s\mu\,\ast} }{m_{S_3}^2}\nonumber\\
&= -\frac{\pi v^2}{\lambda_t \alpha_\mathrm{em}  }  \frac{\sin 2\theta \,\left| y_L^{c\mu }\right|^2  }{2\,m_{S_3}^2}\,.
\end{align}
This closes our discussion about the model proposed in Ref.~\cite{Becirevic:2018afm} and its relation to the couplings $g_{S_L}(m_b)$ and $\delta C_9$.

\section{Phenomenological Analysis - Update}

\subsection{$R_{D^{(\ast)}}$}

 We should first remind the reader that the quantities based on $b\to c\tau \bar \nu$, for which the experimental hints of lepton flavor universality violation (LFUV) have been reported, are defined as,
\begin{equation}
R_{D^{(\ast)}} = \left. \dfrac{\mathcal{B}(B\to D^{(\ast)} \tau\bar{\nu})}{\mathcal{B}(B\to D^{(\ast)} l \bar{\nu})}\right|_{l\in \{e,\mu\}}.
\label{eq:RD_definition}
\end{equation}
\begin{figure}[h]
\includegraphics[scale=0.9]{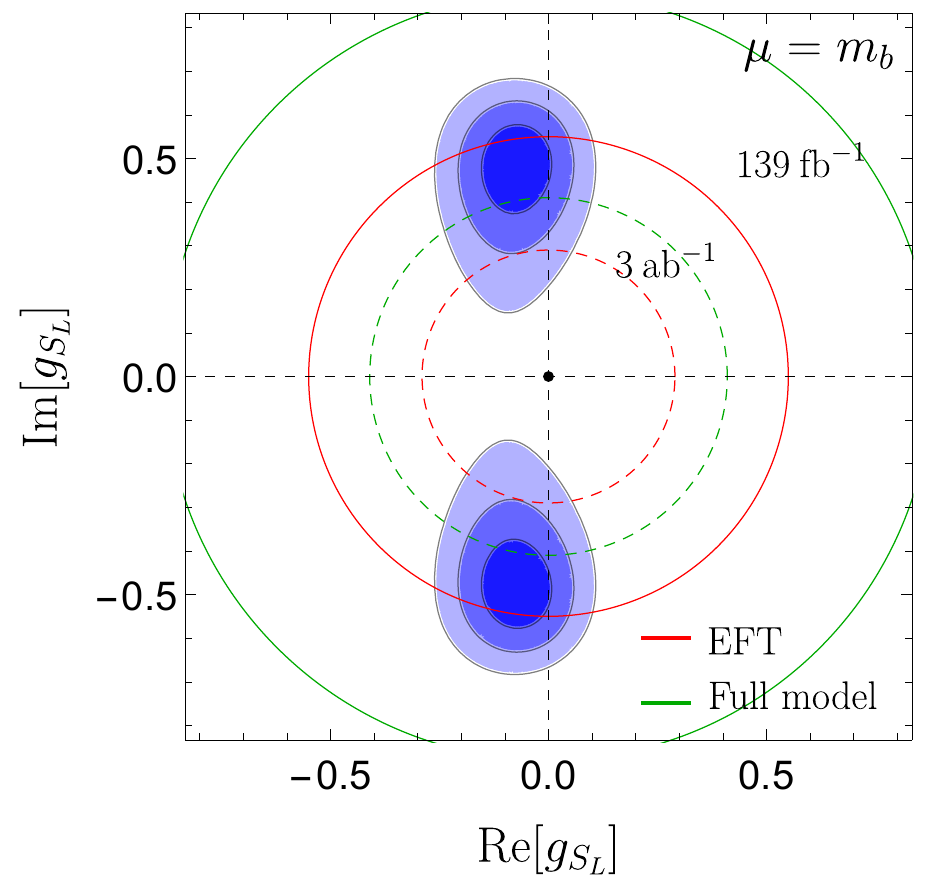}
  \caption{$1\sigma$, $2\sigma$ and $3\sigma$ regions of complex values for $g_{S_L}\equiv g_{S_L}(m_b)$ allowed by $R_{D^{(\ast)}}^\mathrm{exp}$, cf. Eqs.~(\ref{eq:hamiltonian-semilep},\ref{eq:gst}). Red and green circles correspond to the constraints on this coupling obtained from analysis of 
 the high-$p_T$ tail of $pp\to \tau \nu$, as obtained from the LHC data~\cite{ATLAS:2018ihk,ATLAS:2021bjk}. See text for more details. 
  \label{fig:Fpotato}}
\end{figure}
Both quantities have been recently measured~\cite{Belle:2019gij}, and the new experimental averages are~\cite{HFLAV:2019otj}:
\begin{align}
&R_{D}^\mathrm{exp} =0.340\pm 0.030\,, \quad R_{D^\ast}^\mathrm{exp} =0.295\pm 0.014\,,\nonumber \\
& R_D^\mathrm{SM} = 0.293\pm0.008\,,\quad R_{D^\ast}^\mathrm{SM} =0.248\pm 0.001\,,
\end{align}
where we also give the SM predictions in order to emphasize that $R_{D^{(\ast)}}^\mathrm{exp}>R_{D^{(\ast)}}^\mathrm{SM}$, which is often referred to as the $B$-anomalies in charged currents, thus significant to a little over $4 \sigma$. A similar tendency has been observed in a corresponding ratio involving $\cb (B_c\to J/\psi \ell\bar \nu )$~\cite{Aaij:2017tyk}. It should be noted that a similar LFUV effect has been recently tested at LHCb through $\Lambda_b \to \Lambda_c \tau\bar \nu$, and the resulting experimental value $R_{\Lambda_c} = 0.242\pm 0.076$~\cite{LHCb:2022piu}, due to its large uncertainty, is consistent with $R_{\Lambda_c}^\mathrm{SM} =0.333\pm 0.013$, even though it may appear different from what has been observed with decays involving mesons. 
In order to obtain the allowed range of values for $g_{S_L}$ we use the most recent determination of $R_{D^{(\ast)}}^\mathrm{SM}$, obtained after combining the lattice QCD results for the relevant form factors in the high $q^2$-region with those extracted from experimental analysis at low $q^2$'s~\cite{FermilabLattice:2021cdg}. 
Notice also that in this work we use expressions and the values of the ratios of tensor form factors and the (dominant) axial form factor [$A_1(q^2)$] from Ref.~\cite{Bernlochner:2017jka}. A lattice QCD computation of the tensor form factors would be very welcome. From Fig.~\ref{fig:Fpotato} we see that for all currently viable values of $g_{S_L}$, that are consistent with $R_{D^{(\ast)}}^\mathrm{exp}$, one must have $\mathrm{Im}[ g_{S_L} ]\neq 0$. That is why we emphasized after Eq.~\eqref{eq:yL-S3} that one of the couplings entering the expression for $g_{S_L}$ in Eq.~\eqref{eq:gst} should be complex, which we chose to be $y_R^{b\tau}$.

\subsection{Constraints on $g_{S_L}$ from high $p_T$ tails of $pp\to \tau \nu$}

In recent years we witnessed a revival of the idea, first proposed in Ref.~\cite{Eboli:1987vb}, to search for the presence of $\mathcal{O}(\mathrm{TeV})$ LQ's from  the experimental information on the high $p_T$ tails of $pp\to \ell\ell$. That indeed turns out to be a source of interesting constraints on the corresponding Yukawa couplings~\cite{Faroughy:2016osc}. In Ref.~\cite{Greljo:2018tzh} it was proposed to bound the couplings relevant to the explanation of the $R_{D^{(\ast)}}$ anomaly from the high $p_T$ tail of $pp\to \tau\nu$ processes accompanied by the low-$p_T$ jets.  Further details regarding that analysis were provided in Ref.~\cite{Marzocca:2020ueu} where the same set of experimental data were used and the similar bounds on NP couplings obtained.   

More specifically, one is focused on the high-$p_T$ tails in which $\sigma(pp\to \tau^\pm\nu)$ can be written in terms of the partonic cross sections ($\hat \sigma$) and the luminosity functions $\mathcal{L}_{q_i\bar{q}_j}$, summed over all flavors, namely,   
\begin{align}
&\sigma(pp\to \tau^+\nu)=\sum_{ij}\int_0^1\frac{d\tau}{\tau}\mathcal{L}_{q_i\bar{q}_j}(\tau)\left[\hat{\sigma}(\tau s)\right]_{ij},\\
&\mathcal{L}_{q_i\bar{q}_j}=\tau \int_y^1\frac{dx}{x}\bigl(f_{q_i}(x,\mu_F)f_{\bar{q}_j}(\tau/x, \mu_F)+q_i\leftrightarrow \bar{q}_j\bigr),
\end{align}
where the known partonic distribution functions $f_{q_i}(x,\mu_F)$ depend on the factorization scale $\mu_F$, that is conveniently taken to be the partonic center of mass energy $\sqrt{\hat{s}}$.  Note also that the kinematic variable $\tau = \hat s/s = m_{\tau^\pm \nu}^2/s$.

The channel of interest for our discussion is $c\bar{b}\to\tau^+\nu_\tau$, that in Refs.~\cite{Greljo:2018tzh,Marzocca:2020ueu} was assumed to be described/parametrized by the low energy effective theory~\eqref{eq:hamiltonian-semilep} which then contributes to the partonic cross via,
\begin{align}
\hat{\sigma}(\hat{s})&=\frac{|V_{cb}|^2 G_F^2 \hat{s}}{18\pi}\left( \frac{3}{4}|g_{S_{L}}|^2+ 4|g_T|^2\right) ,
\end{align}
in addition to the SM contribution, which in this analysis is treated as a background. In our case the two couplings are related, cf. Eq.~\eqref{eq:gst}. 
After recasting to our problem the bounds on $W^\prime$ obtained from $139~\mathrm{fb}^{-1}$ by ATLAS~\cite{ATLAS:2021bjk}, we get the bound on $g_{S_L} = 4 g_T$, which upon evolving down to $\mu=m_b$ amounts to 
\bea\label{boundEFT}
\left| g_{S_L} \right| \leq 0.51\,,
\eea  
represented by the red circle in Fig.~\ref{fig:Fpotato}.

Since the LQ masses that we work with are not far away from the high-$p_T$ tails accessed in experiments, one should also use the propagating LQ, and check on the difference with respect to the bounds on $g_{S_L}$ obtained by treating LQ as static. Using the Lagrangian specified in Sec.~\ref{sec:1}, for the partonic cross section we obtain:   
\begin{align}
&\frac{{\rm d}\hat{\sigma}\left(c\bar{b}\to\tau^+\nu_\tau\right)}{{\rm d}\hat{t}} = \frac{1}{192\pi \hat{s}^2}\bigg[ \, \frac{g^4\left|V_{cb}\right|^2\hat{t}^2}{\left(\hat{s}-m_W^2\right)^2}  \biggr. \nonumber \\
& + 
\frac{\sin^2\theta\left|V_{cs}\cos\theta-V_{cb}\sin\theta\right|^2\left|y^L_{c\tau}\right|^4\hat{t}^2}{4\left(\hat{t}-m_{S_3}^2\right)^2} 
+\frac{\left|y^L_{c\tau}\right|^2\left|y^R_{b\tau}\right|^2\hat{u}^2}{\left(\hat{u}-m_{R_2}^2\right)^2}\nonumber \\
&+\frac{g^2\sin\theta\:{\rm Re}\left[\left(V_{cs}\cos\theta-V_{cb}\sin\theta\right)V_{cb}^*\right]\left|y^L_{c\tau}\right|^2\hat{t}^2}{\left(\hat{s}-m_W^2\right)\left(\hat{t}-m_{S_3}^2\right)}\bigg],
\end{align}
with a similar expression for $b\bar{c}\to\tau^-\bar \nu_\tau$, 
where the first term within the brackets corresponds to the SM contribution, followed by the $S_3$ and $R_2$ contributions, and finally the last term is interference between $S_3$ and the SM contributions. Note that the fermion masses in the above expression have been neglected. 
It appears that, for our phenomenological application, the $R_2$ term indeed dominates because the $S_3$ term is suppressed with respect to $R_2$ by $V_{cs}\cos\theta-V_{cb}\sin\theta$, in which the first term is small due to a tiny $\cos\theta$ and the second one due to the smallness of $V_{cb}$.  One can therefore write: 
\begin{align}
\hat{\sigma}(\hat{s})&\simeq\frac{|y^R_{b\tau}|^2\left(|y^L_{c\tau}|^2+|y^L_{c\mu}|^2\right)}{192\pi m_{R_2}^2}\left[\frac{x+2}{x(1+x)}-\frac{2\log(1+x)}{x^2}\right],
\label{eq:FullPartonic}
\end{align}
where $x=\hat{s}/m_{R_2}^2$. Again, after recasting the results by ATLAS~\cite{ATLAS:2021bjk} and using the above expressions, we obtain 
\bea
\left(\left|y^L_{c\tau}\right|^2+\left|y^L_{c\mu}\right|^2\right)\left|y^R_{b\tau}\right|^2 < 5.95\,,
\eea
which then can be combined into $g_{S_L}$ via Eq.~\eqref{eq:gst}, and evolved down to $\mu=m_b$. For the benchmark mass, $m_{R_2}=1.3~\mathrm{TeV}$, we then find,
\bea
\left| g_{S_L} \right| \leq 0.88\,,
\eea  
shown by a green circle in Fig.~\ref{fig:Fpotato}. Note that this bound, obtained by including the propagating $R_2$, is far less stringent than the one deduced from the data after integrating out $R_2$, c.f. Eq.~\eqref{boundEFT}. Of course, if  the LQ is taken to be heavier, such as $m_{R_2}\gtrsim 5$~TeV, the bounds obtained from the effective theory would be much closer to the one in which the propagating LQ is considered, cf. also Ref.~\cite{Iguro:2020keo}. More details on this analysis and the notation employed above can be found in Ref.~\cite{Jaffredo:2021ymt}.

In summary, from the current data by ATLAS regarding the mono-tau high-$p_T$ tails, and by including the propagation of the $R_2$ LQ of $m_{R_2}=1.3~\mathrm{TeV}$, 
one cannot obtain a very useful constraint on the NP couplings appearing in Eq.~\eqref{eq:hamiltonian-semilep}. However, by assuming data to be Gaussianly distributed, 
one can make a simple projection to an integrated $3~\mathrm{ab}^{-1}$ of the LHC data and arrive at $\left| g_{S_L} \right| \leq 0.41$, which would indeed be a powerful 
constraint. In Fig.~\ref{fig:Fpotato} the dashed circles correspond to the projected bounds both by using the effective and propagating $R_2$ of $m_{R_2}=1.3$~TeV.

\subsection{$R_{K^{(\ast)}}$}

As for the $b\to s\ell\ell$ decays, the LFUV ratios are~\cite{Hiller:2003js} 
\begin{equation}
R_{K^{(\ast)}}^{[q_1^2, q_2^2]} =  \dfrac{\mathcal{B}'(B\to K^{(\ast)} \mu\mu)}{\mathcal{B}'(B\to K^{(\ast)} ee)}  \,,
\label{eq:RK_definition}
\end{equation}
where $\cb^\prime$ stands for the partial branching fraction taken over the common interval, $[q_1^2, q_2^2]$, conveniently chosen as  
to stay away from the region in which the $\bar cc$-resonances dominate the di-lepton spectra. We include the most recent value for $R_K^{[1.1,6]}$~\cite{LHCb:2021trn}, and  for completeness we also quote $R_{K^\ast}^{[1.1,6]}$~\cite{Aaij:2017vbb}: 
\begin{equation}
R_{K}^\mathrm{exp} =0.847 \pm 0.042\,, \quad R_{K^\ast}^\mathrm{exp} =0.71 \pm 0.10\,.
\end{equation}
These values are smaller than predicted in the SM, $R_{K^{(\ast)}}^{[1,6]} = 1.00(1)$~\cite{Bordone:2016gaq}. This apparent LFUV, 
$R_{K^{(\ast)}}^\mathrm{exp}<R_{K^{(\ast)}}^\mathrm{SM}$, is attributed to a deficit of the muon pairs in the final state with respect to the SM prediction. 
Another novel  $b\to s\mu\mu$ result is the most recent LHCb measurement, $\mathcal{B}(B_s\to\mu\mu)= (3.09^{+0.48}_{-0.44}) \times 10^{-9} $~\cite{LHCb:2021awg}, 
which after combining with the other two LHC experiments leads to,
\begin{equation}
\mathcal{B}(B_s\to\mu\mu)= \left. 2.85(33) \times 10^{-9}\right|_\mathrm{exp}  \! ,  \left. 3.66(14)\times 10^{-9}\right|_\mathrm{SM},
\end{equation}
showing that the measured value is about $2\sigma$ {\it smaller} than predicted in the SM~\cite{Beneke:2019slt}.
These three quantities [$R_{K}$, $R_{K^\ast}$, $\mathcal{B}(B_s\to\mu\mu)$] are then used to determine~\cite{Angelescu:2018tyl} 
\bea
\delta C_{9}=-\delta C_{10} = - 0.41\pm 0.09\,,
\eea
also consistent with the global fit analyses~\cite{global}. 
To make this result consistent with Eq.~\eqref{eq:sin2theta}, and knowing that $\lambda_t<0$, one should have $\sin2\theta < 0$.  The factor $\sin2\theta$ provides a desired suppression of the $b\to s\mu\mu$ decays with respect to $b\to c\tau\bar \nu$. Indeed, from the fit with data we obtain $|\theta | \approx \pi/2$, i.e. slightly larger than but close to $\pm \pi/2$. The contribution of this model to the $B_s-\bar{B}_s$ mixing amplitude comes from the $S_3$
box-diagram and it is proportional to $\sin^2(2\theta)$, thus again bringing a desired suppression since we know that the SM contribution saturates the measured 
$\Delta m_{B_s}$. More precisely, the $S_3$ contribution to $\Delta m_{B_s}$ is $\propto \sin^2 2\theta \left[\left( y_L^{c\mu}\right)^2 + \left( y_L^{c\tau}\right)^2\right]^2/m_{S_3}^2$.

\subsection{Updating the parameter space of our $R_2$-$S_3$ model}

Besides $R_{D^{(\ast)}}$, $R_{K^{(\ast)}}$ and $\cb(B_s\to \mu\mu)$ discussed so far in this section, and which are the most important constraints on the parameters of this model, the following quantities are 
used as further constraints: 
\begin{figure}[!h]
  \centering
  \includegraphics[scale=0.4]{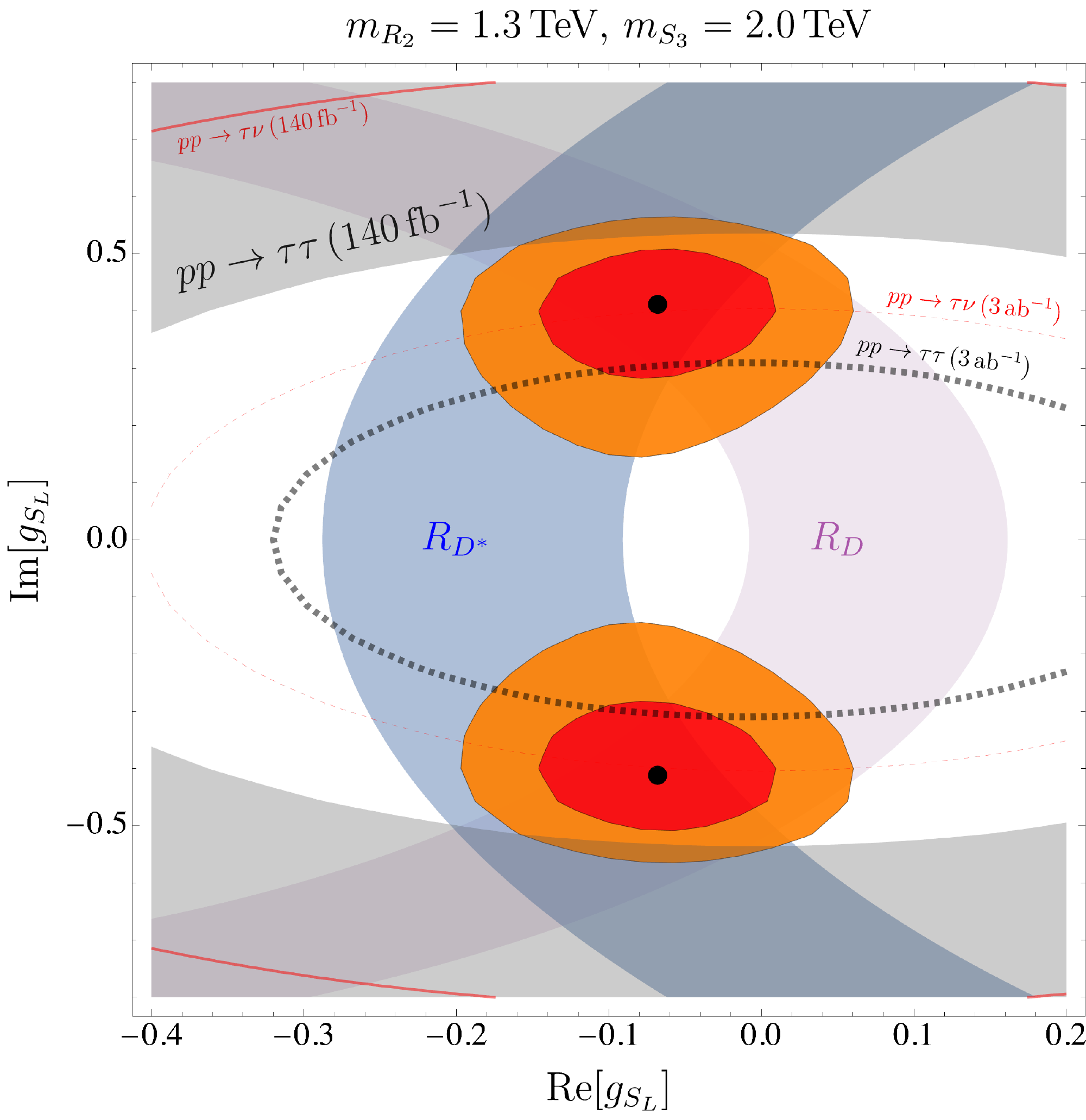}
  \caption{Results of the flavor fit in the $g_{S_L}$ plane, as defined in Eq.~(\ref{eq:hamiltonian-semilep}) for the transition $b\to c\tau \bar{\nu}_\tau$. The allowed $1\,\sigma (2\,\sigma)$ regions are shown in red (orange). Separate constraints from $R_D$ and $R_{D^\ast}$ to $2\,\sigma$ accuracy are shown by the purple and blue regions, respectively. The current LHC exclusions are depicted by the gray regions. We also show the projected bounds expected to be obtained from the high $p_T$ mono-tau (red curve) and di-tau tails (dashed curve) 
 with $3~\mathrm{ab}^{-1}$ of data. 
  \label{fig:gSplotRDst}}
\end{figure}
\begin{itemize}
\item The $B_s-\bar{B}_s$ mixing is included by considering $R(\Delta m_{B_s}) = \Delta m_{B_s}/\Delta m_{B_s}^\mathrm{SM}$. We combine the improved experimental value with the  lattice QCD result by HPQCD~\cite{Dowdall:2019bea} and obtain   
$R(\Delta m_{B_s})^\mathrm{exp}  = 1.027(68)$. We also performed the full scan of parameter space using the FLAG value for the corresponding hadronic matrix element as computed with $N_f=2+1$ dynamical quark flavors, which corresponds to $R(\Delta m_{B_s})^\mathrm{exp}  = 0.897(69)$~\cite{Aoki:2019cca}, and found no significant impact to the selected parameter space, except for the slightly different value of $\chi^2_\mathrm{min}$. 

\item We require the results to be consistent with $R_{D^{(\ast )}}^{(\mu/e)\,\mathrm{exp}}= 0.977(43)$~\cite{Dorsner:2017ufx}, which is obtained by combining $R_{D}^{(\mu/e)\,\mathrm{exp}}=0.995(45)$~\cite{Glattauer:2015teq} with $R_{D^\ast}^{(\mu/e)\,\mathrm{exp}}=1.04(5)$~\cite{Abdesselam:2017kjf}. Note that in this model only $S_3$ can contribute to $R_{D^{(\ast )}}^{(\mu/e)} = \mathcal{B}(B\to D^{(\ast )}\mu \bar \nu)/  \mathcal{B}(B\to D^{(\ast )}e \bar \nu)$.

\item We also impose the measured $\mathcal{B}(B\to \tau \nu) = 1.09(24)\times 10^{-4}$~\cite{ParticleDataGroup:2020ssz} as a constraint, where we use $f_B=190.0\pm 1.3$~MeV~\cite{Aoki:2019cca}.  When needed we take the CKM couplings from Ref.~\cite{Charles:2004jd}.

\item Tests of LFUV in the kaon leptonic decays can also be used as constraints to the $S_3$ LQ. We consider 
$r_K^{(e/\mu)}
= \Gamma (K\to e \bar \nu)/ \Gamma(K\to \mu \bar \nu)$ and 
$r_K^{(\tau/\mu)}= \Gamma (\tau \to K \bar \nu)/ \Gamma(K\to \mu \bar \nu)$, the measured values of which~\cite{ParticleDataGroup:2020ssz} are compared to the SM values, $r_K^{(e/\mu)\,\mathrm{exp} }/r_K^{(e/\mu)\,\mathrm{SM} } =1.004(4)$, $r_K^{(\tau/\mu)\,\mathrm{exp} }/r_K^{(\tau/\mu)\,\mathrm{SM} } =0.972(14)$,  and represent a rather powerful constraint, cf. Ref.~\cite{Dorsner:2017ufx}. Similarly, the ratio $r_{D_s}^{(\tau/\mu)}= \cb(D_s\to \tau \bar\nu)/\cb(D_s\to \mu\bar \nu)$, is converted to a constraint when comparing to $r_{D_s}^{(\tau/\mu)\,\mathrm{exp} }/r_{D_s}^{(\tau/\mu)\,\mathrm{SM} } =1.027(52)$~\cite{ParticleDataGroup:2020ssz}. 

\item Experimental bounds on the lepton flavor violating (LFV) decay modes $\mathcal{B}(\tau \to \mu\gamma )< 4.4 \times 10^{-8}$, $\mathcal{B}(\tau \to \phi \mu)<8.4 \times 10^{-8}$~\cite{ParticleDataGroup:2020ssz} provide the significant constraints too. Note that both $R_2$ and $S_3$ contribute to the latter mode, cf. Ref.~\cite{Dorsner:2016wpm}, while the expression for the LQ contribution to $\mathcal{B}(\tau \to \phi \mu )$ can be found in~\cite{Becirevic:2016oho}. We also use $\cb(B \to K\mu^- \tau^+)< 2.8\times 10^{-5}$~\cite{ParticleDataGroup:2020ssz,LHCb:2020khb} in our scan of the parameter space.

\item In Ref.~\cite{Becirevic:2018afm} we provided the expressions for $R_{\nu\nu}^{(\ast)}=\cb(B\to K^{(\ast )}\nu\bar \nu)/\cb(B\to K^{(\ast )}\nu\bar \nu)^\mathrm{SM}$, which should respect the experimental bounds $R_{\nu\nu}<3.9$ and $R_{\nu\nu}^{\ast}<2.7$~\cite{Grygier:2017tzo}.

\item The complete expressions for for the scalar LQ contributions to $\cb (Z\to \ell \ell)$ have been derived in Ref.~\cite{Arnan:2019olv} and they are used in this analysis, together with the experimental values for the branching fractions given in Ref.~\cite{ParticleDataGroup:2020ssz}. 

\item Finally, we take into account the bounds on the couplings derived from the high-$p_T$ tails after recasting the bounds on heavy Higgs decaying to two $\tau$-leptons obtained from $139~\mathrm{fb}^{-1}$ of data by ATLAS, reported in Ref.~\cite{ATLAS:2020zms}. By focusing on the region of $m_{\tau\tau}\geq 700$~GeV, and by using the propagating $R_2$ of $m_{R_2}=1.3$~TeV, we obtain rather stringent bounds on the couplings, which can be conveniently written as
\begin{align}
& 1.75 (y_R^{b\tau})^4 + 0.29  (y_R^{b\tau})^2 
 + 7.96   (y_L^{c\tau})^4 + 3.43  (y_L^{c\tau})^2 \leq 25.9 .
\end{align} 
Notice that in obtaining this result we use the experimental bounds from Ref.~\cite{ATLAS:2020zms} to $2\sigma$.
\end{itemize}

A careful reader would notice that with respect to Ref.~\cite{Becirevic:2018afm}, where $m_{R_2}=0.8$~TeV has been used to present the results, here we take $m_{R_2}=1.3$~TeV. This choice is made in order to be consistent with the most recent bounds regarding the LQ production processes in the direct searches at the LHC, as discussed in Ref.~\cite{Angelescu:2018tyl}. For the same reason we take $m_{S_3}=2$~TeV and perform a scan over the remaining parameters of the model, $y_R^{b\tau}$, $y_L^{c\mu}$, $y_L^{c\tau}$ and $\theta \in (\pi/2,\pi)\cup (-\pi/2,0)$, by imposing all of the constraints discussed so far. In Fig.~\ref{fig:gSplotRDst} we show the result of such a scan in the $g_{S_L}\equiv g_{S_L}(m_b)$ plane. We obtain $\chi^2_\mathrm{min}=13.5$, and for the best fit values we get (to $1\sigma$)
\begin{equation}
\label{eq:bfv}
  \mathrm{Re} [g_{S_L}] = -0.07(14),~
  |\mathrm{Im} [g_{S_L}]| = 0.44
     \left(^{+0.09}_{-0.12}\right) \,.
\end{equation}
If we did not use the experimental bounds on the LFV modes as constraints, our flavor fit would have given two solutions: one corresponding to a small angle $\theta \sim 0$, and another one corresponding to $|\theta| \sim \pi/2$. In fact, $\mathcal{B}(\tau\to\mu\phi)\propto \cos^4\theta$, and the corresponding experimental bound help us select a viable solution, i.e. the one with $|\theta| \approx \pi/2$. 
In Fig.~\ref{fig:gSplotRDst} we also plot the current constraint, $\vert g_{S_L}\vert < 0.55$, obtained from the study of the high-$p_T$ di-tau tails. In the same plot we also show the projected bound from $3~\mathrm{ab}^{-1}$ of data, a constraint which on the basis of current information should be much stronger than the one based on the high-$p_T$ mono-tau tails.

Before closing this Section we also provide the ranges for the couplings we obtain after imposing all of the constraints discussed so far:
\begin{align}
\label{eq:Yvalues}
y_L^{c\mu}&\in (0.16,0.33)_{1\sigma},  (0.11,0.40)_{2\sigma},\nonumber\\
y_L^{c\tau}&\in (0.87,1.40)_{1\sigma},  (0.64,1.54)_{2\sigma},\nonumber\\
\mathrm{Re}\left[ y_R^{b\tau}\right] &\in (-0.37,0.02)_{1\sigma},  (-0.58,0.15)_{2\sigma},\nonumber\\
\left| \mathrm{Im}\left[ y_R^{b\tau}\right] \right| &\in (0.83,1.53)_{1\sigma},  (0.61,1.87)_{2\sigma},\nonumber\\
\theta &\in \frac{\pi}{2}(1.01,1.06)_{1\sigma}, \frac{\pi}{2} (1.01,1.12)_{2\sigma},
\end{align}
where  $\mathrm{Im}\left[ y_R^{b\tau}\right]$ has two symmetric solutions (positive and negative). 

\section{More Observables}

\subsection{Contribution to the electric dipole moment of the neutron}
From fit to the data we saw that we obtain $\mathrm{Im}[ g_{S_L}] \gg \mathrm{Re}[ g_{S_L}]$ 
when accommodating $R_{D^{(\ast )}}^\mathrm{exp}>R_{D^{(\ast )}}^\mathrm{SM}$. In other words we get a large $|\mathrm{Im}\left[ y_R^{b\tau}\right]|$, which then calls for a careful analysis 
of the observables in which such a complex phase may play a significant role. 
We first check whether or not this phase might be in conflict with the current bound on the electric 
dipole moment of the neutron, $|d_n| < 1.8 \times 10^{-26}$~$e$cm~\cite{Abel:2020gbr}. That issue has recently been addressed in Ref.~\cite{Dekens:2018bci} in the scenarios in which the SM is extended by one or more scalar leptoquarks. For our purpose it is important to note that the charm quark contribution to $d_n$ can be written as 
$d_n = g_T^c \, d_c$, where the tensor charge $g_T^c$, defined as 
\bea
\langle N\vert \bar c\sigma^{\mu\nu} \gamma_5 c\vert N\rangle = g_T^c \, \bar u_N \sigma^{\mu\nu} \gamma_5  u_N \,,
\eea
has been recently computed by means of numerical simulations of QCD on the lattice with $N_{\mathrm{f}}=2+1+1$ dynamical quark flavors~\cite{Alexandrou:2019brg}. The reported  result at $\mu=2$~GeV, in the $\overline{\mathrm{MS}}$ renormalization scheme is $g_T^c=-(2.4\pm 1.6)\times 10^{-4}$. We translate the notation of Ref.~\cite{Dekens:2018bci} to the one used here and write:
\bea
d_c& =&  0.1 \times Q_c \, e \, m_c\,  {1\over m_{R_2}^2}\, \mathrm{Im}\left[ V_{cb}^\ast \, y_R^{b\tau\,\ast}  \, y_L^{c\tau } \right]\nonumber\\
& \simeq &  0.1 \times Q_c \, e \, m_c\,  {4\sqrt{2} G_F V_{cb}^2 \over 1.7}\, \mathrm{Im}\left[ g_{S_L} \right],
\eea
where in the second line we employed Eq.~\eqref{eq:gst}. In the denominator $1.7$ accounts for the running of $g_{S_L}$ to the low energy scale. By using the charm quark mass value from Ref.~\cite{Aoki:2019cca}, the central value for $g_T^c$, and the experimental bound on $|d_n| $, we arrive at 
\bea
\vert \mathrm{Im}\left[ g_{S_L} \right] \vert < 0.76\,,
\eea
which is obviously in good agreement with what we obtain in Fig.~\ref{fig:gSplotRDst} and in Eq.~\eqref{eq:bfv}. However, we should note that if instead of the central value we take $g_T^c = -4\times 10^{-4}$ then this constraint translates to $\vert \mathrm{Im}\left[ g_{S_L} \right] \vert < 0.46$, which would eliminate a fraction of the allowed $g_{S_L}$ regions in Fig.~\ref{fig:gSplotRDst}. This shows why a more precise lattice QCD value of $g_T^c$ would be highly beneficial for checking the validity of the model proposed in Ref.~\cite{Becirevic:2018afm} and further discussed here.

\subsection{Contribution to $\Delta a_\mathrm{CP}$}

The difference in the time-integrated CP asymmetries of $D^0\to K^+K^-$ and $D^0\to \pi^+\pi^-$ has been measured by LHCb. Their recent result 
$\Delta A_\mathrm{CP} = (-15.4\pm 2.9) \times 10^{-4}$~\cite{Aaij:2019kcg} has been corrected for the effects of $D^0-\overline D^0$ mixing so that the result for the difference of  direct CP asymmetries becomes $\Delta a_\mathrm{CP}^\mathrm{dir} = (-15.7\pm 2.9) \times 10^{-4}$~\cite{Lenz:2020awd}. The interpretation of this result is still unclear. In the SM picture the effect could be attributed to the (nonperturbative) rescattering of light mesons in the final state. Otherwise, one would need a NP contribution to accommodate the measured value~\cite{Grossman:2019xcj}.

In Ref.~\cite{Giudice:2012qq} the NP contribution to $\Delta a_\mathrm{CP}$ has been estimated under the assumption of the maximal strong phases. It was found that $|\Delta a_\mathrm{CP}| \lesssim 1.8|\mathrm{Im} C_8^{\mathrm{NP}}(m_c)+\mathrm{Im}C_{8^\prime}^{\mathrm{NP}}(m_c)|$, where $C_{8,8'}$ are the Wilson coefficients of the chromomagnetic operators:
\begin{equation}
  \label{eq:Chromos}
  {\cal H} = \frac{G_F}{\sqrt{2}}\,\frac{g_s m_c}{4\pi^2}\, \bar u_L \sigma_{\mu\nu} \left[C_8 P_R + C_{8'} P_L \right] c\ T^a G_a^{\mu\nu}.
\end{equation}
In our model $R_2$ will contribute to $c_R \to u_L g$ and to one-loop we get 
\begin{equation}
C_8^{\mathrm{NP}} = \frac{m_\tau V_{ub} y_R^{b\tau}{y_L^{c\tau}}^\ast}{4 \sqrt{2} G_F m_c} B_0^\prime(0, m_{R_2}^2, m_\tau^2) \,.
\end{equation}
With the structure of couplings chosen in our model, cf. Eq.~\eqref{eq:yL-yR}, there is no one-loop contribution to $c_L \to u_R g$, i.e. $C_{8'} =0$. 
By taking $m_\tau /m_{R_2}\to 0$, we have $B_0^\prime(0, m_{R_2}^2, 0) \to 1/(2 m_{R_2}^2)$, which then leads to $|\Delta a_{CP}|  \lesssim 10^{-4}$, thus a very small effect.

\subsection{Contribution to $B\to K \nu\nu$ and $K\to \pi \nu\nu$}

It is well known that a contribution of the left-handed current to $b\to s \ell\ell$ implies a similar contribution 
to $B\to K^{(\ast)}\nu\nu$ decays. In our case that means 
\begin{equation}
\label{eq:Rnunu}
R_{\nu\nu}^{(\ast)} ={\mathcal{B}(B\to K^{(\ast)}\nu\nu)\over \mathcal{B}(B\to K^{(\ast)}\nu\nu)^{\mathrm{SM}}}=  \dfrac{\sum_{ij}|\delta_{ij}C_L^{SM}+\delta C_{L}^{ij}|^2}{3|C_L^{\mathrm{SM}}|^2},
\end{equation}
where $C_{L}^{\mathrm{SM}}=-6.38(6)$~\cite{Altmannshofer:2009ma} and the tree-level contribution arising from $S_3$ amounts to~\cite{Dorsner:2017ufx} 
\begin{align}
\sum_{ii} \delta C_L^{ii} &= \sum_{i} \dfrac{\pi v^2}{2\alpha_{\mathrm{em}} \lambda_t}\frac{ \left( Y_L^{(S_3)}\right)^{b\, i } \left( Y_L^{(S_3)}\right)^{s\, i\,\ast} }{m_{S_3}^2},\nonumber\\
& = -  \dfrac{\pi v^2}{2\alpha_{\mathrm{em}} \lambda_t}\frac{ \sin 2\theta \,  \left( y_L^{c\mu\,^2}+ y_L^{c\tau\,^2} \right) }{m_{S_3}^2}\,,
\end{align}
thus also negative, and therefore the net effect in the present model is that $R_{\nu\nu}^{(\ast)} > 1$. We get  
\bea\label{eq:nunu}
R_{\nu\nu}^{(\ast)}\in (1.3, 2.5)_{1\sigma}, (1.1,3.4)_{2\sigma},
\eea
the result which is likely to be probed experimentally at Belle~II~\cite{Dattola:2021cmw}.

The expressions relevant to the $S_3$ contribution to $\cb(K\to \pi \nu\nu)$ have been derived in Ref.~\cite{Fajfer:2018bfj}. With our choice of couplings, together with values given in Eq.~\eqref{eq:Yvalues}, that contribution turns out to be very small. We checked that the same conclusion holds true for the $R_2$ contribution as well.

\subsection{ $B\to K \mu\tau$ and its correlation with $\tau \to \mu \gamma$ and $R_{\nu\nu}^{(\ast ) }$}
\begin{figure*}[!t]
  \centering
  \includegraphics[scale=0.35]{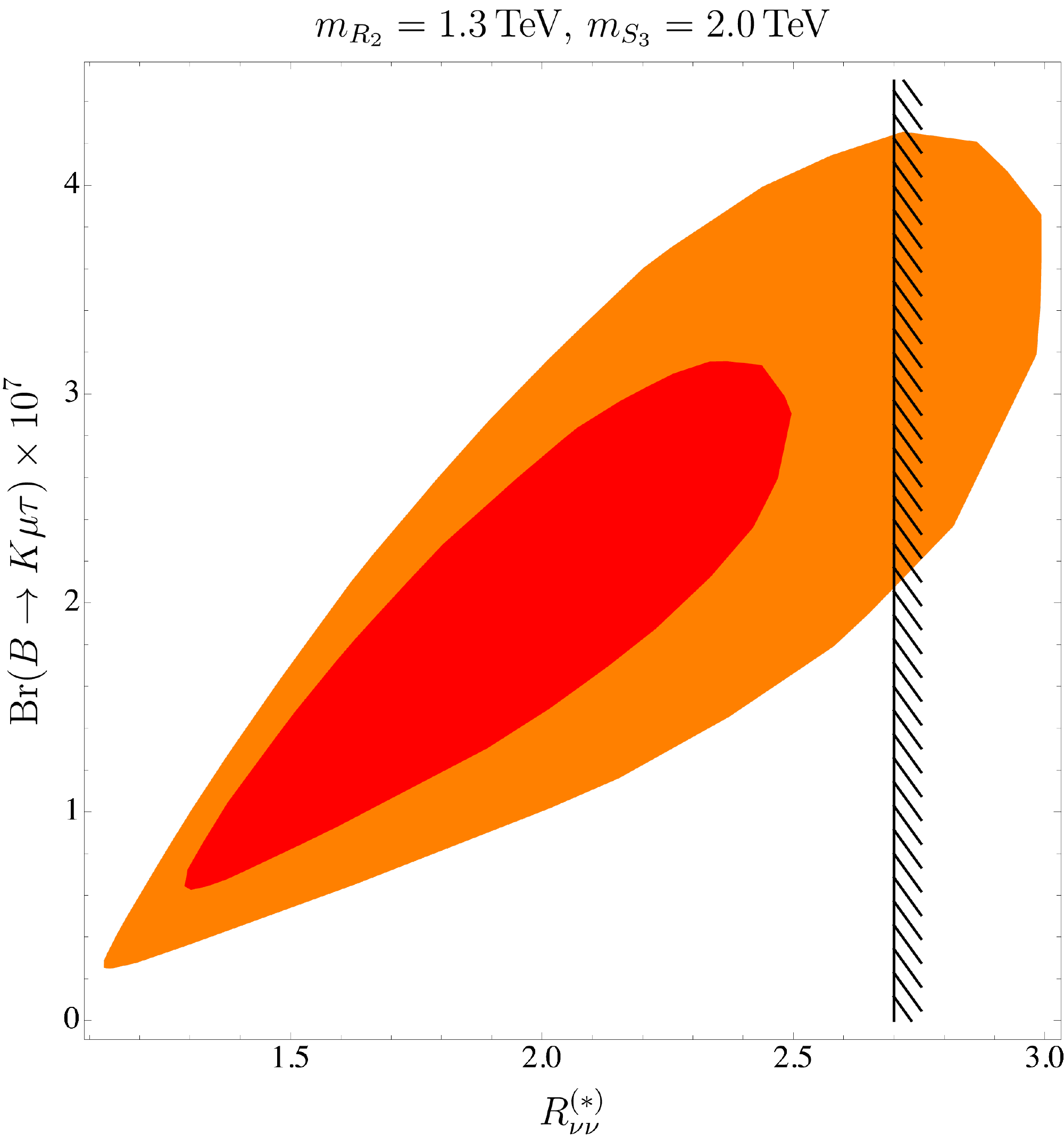}\qquad\includegraphics[scale=0.35]{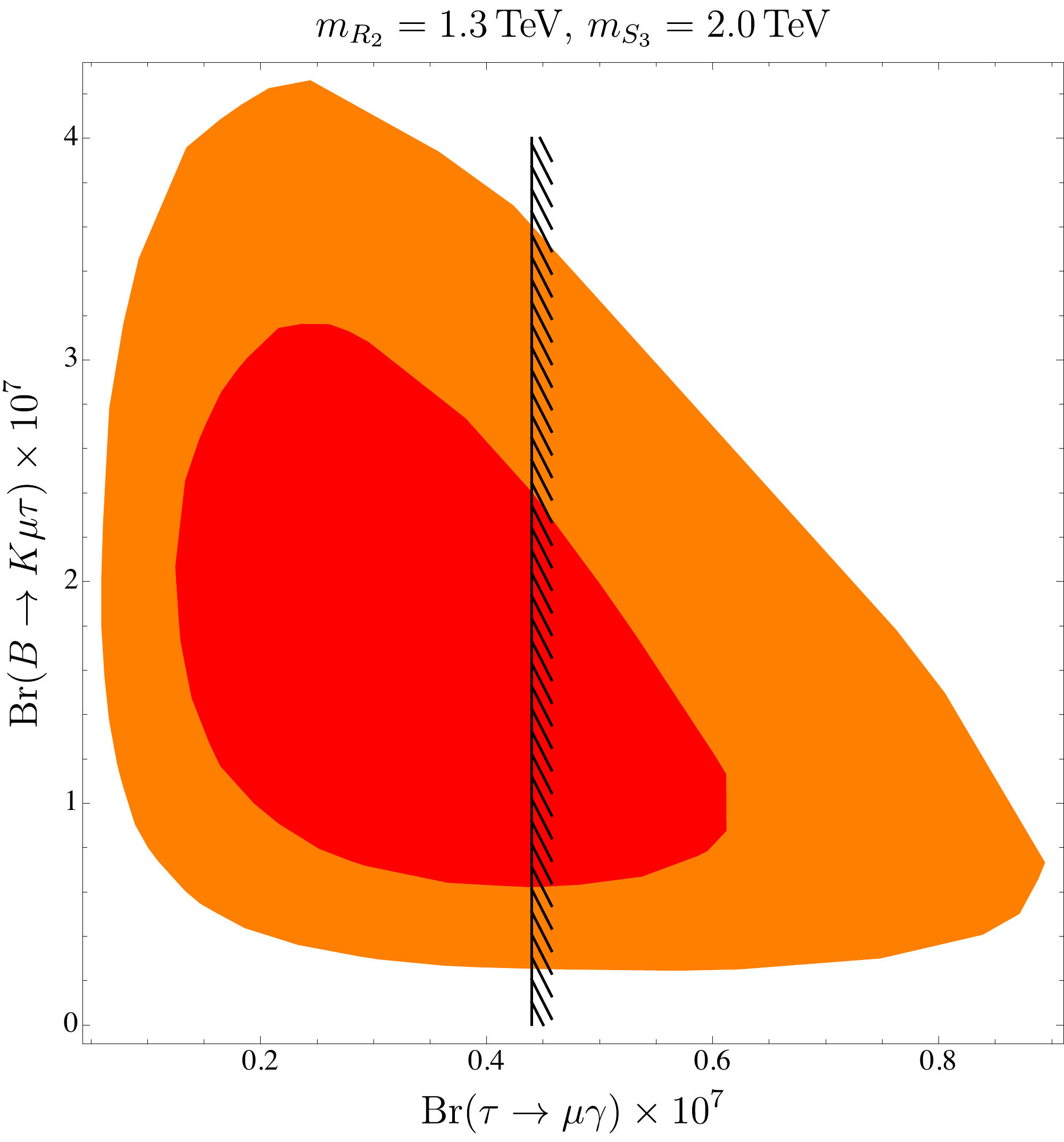}
  \caption{$\mathcal{B}(B\to K \mu \tau)$ is plotted against $R_{\nu\nu}^{(\ast)}=\mathcal{B}(B\to K^{(\ast)} \nu \bar{\nu})/\mathcal{B}(B\to K^{(\ast)} \nu \bar{\nu})^{\mathrm{SM}}$ for the $1\,\sigma$ (red) and $2\,\sigma$ (orange) regions of Fig.~\ref{fig:gSplotRDst}. The black line denotes the current experimental limit, $R_{\nu\nu}^{\ast}<2.7$~\cite{Grygier:2017tzo}. We also show the similar correlation between $\mathcal{B}(B\to K \mu \tau)$ and $\mathcal{B}(\tau\to \mu \gamma)$ obtained in this model.
  \label{fig:BKmutau}}
\end{figure*}

Most of the models that can accommodate the LFUV also predict a non-zero branching fraction of the associated LFV decay modes~\cite{Glashow:2014iga}. 
Even more interesting is that in our model we get both the lower and the upper bounds, namely and to $1\sigma$,
 \begin{equation}
0.6 \times 10^{-7}\lesssim  \mathcal{B}( B\to K \mu^\pm \tau^\mp) \lesssim 3.1\times 10^{-7} \,,
\end{equation}
currently, however, two orders of magnitude lower than the experimental limit~\cite{LHCb:2020khb}.  
This prediction can be translated into similar modes via relations $\mathcal{B}(B\to K^\ast \mu\tau)\approx 1.9\times \mathcal{B}(B\to K \mu \tau)$,  $\mathcal{B}(B_s\to \mu\tau)\approx  0.9 \times \mathcal{B}(B\to K \mu \tau)$, and $\mathcal{B}(\Lambda_b\to \Lambda \mu\tau)\approx 1.7\times \mathcal{B}(B\to K \mu \tau)$~\cite{Becirevic:2016zri}. 
It is interesting to note that $\mathcal{B}(B\to K  \mu\tau)$ is linearly correlated with $R_{\nu\nu}^{(\ast)}$, as show in Fig.~\ref{fig:BKmutau}.

Another interesting LFV mode is $\tau\to \mu \gamma$, because in order to accommodate both types of $B$-anomalies we needed to switch on the NP couplings to both $\tau$ and $\mu$. Indeed, in this model we obtain a lower bound which to $1\sigma$ is
\begin{equation}
 \mathcal{B}(\tau \to \mu \gamma) \gtrsim 1.2\E{-8}\,,
\end{equation}
and its correlation with $\mathcal{B}(B\to K  \mu\tau)$, also shown in Fig.~\ref{fig:BKmutau}, is less pronounced than the one between  $\mathcal{B}(B\to K  \mu\tau)$ and $R_{\nu\nu}^{(\ast)}$.

\subsection{Angular observables in $B\to D^\ast (\to D\pi) \tau\nu$ and in $\Lambda_b\to \Lambda_c(\to \Lambda \pi) \tau\nu$}

\begin{figure}[!h]
  \centering
  \includegraphics[scale=0.14]{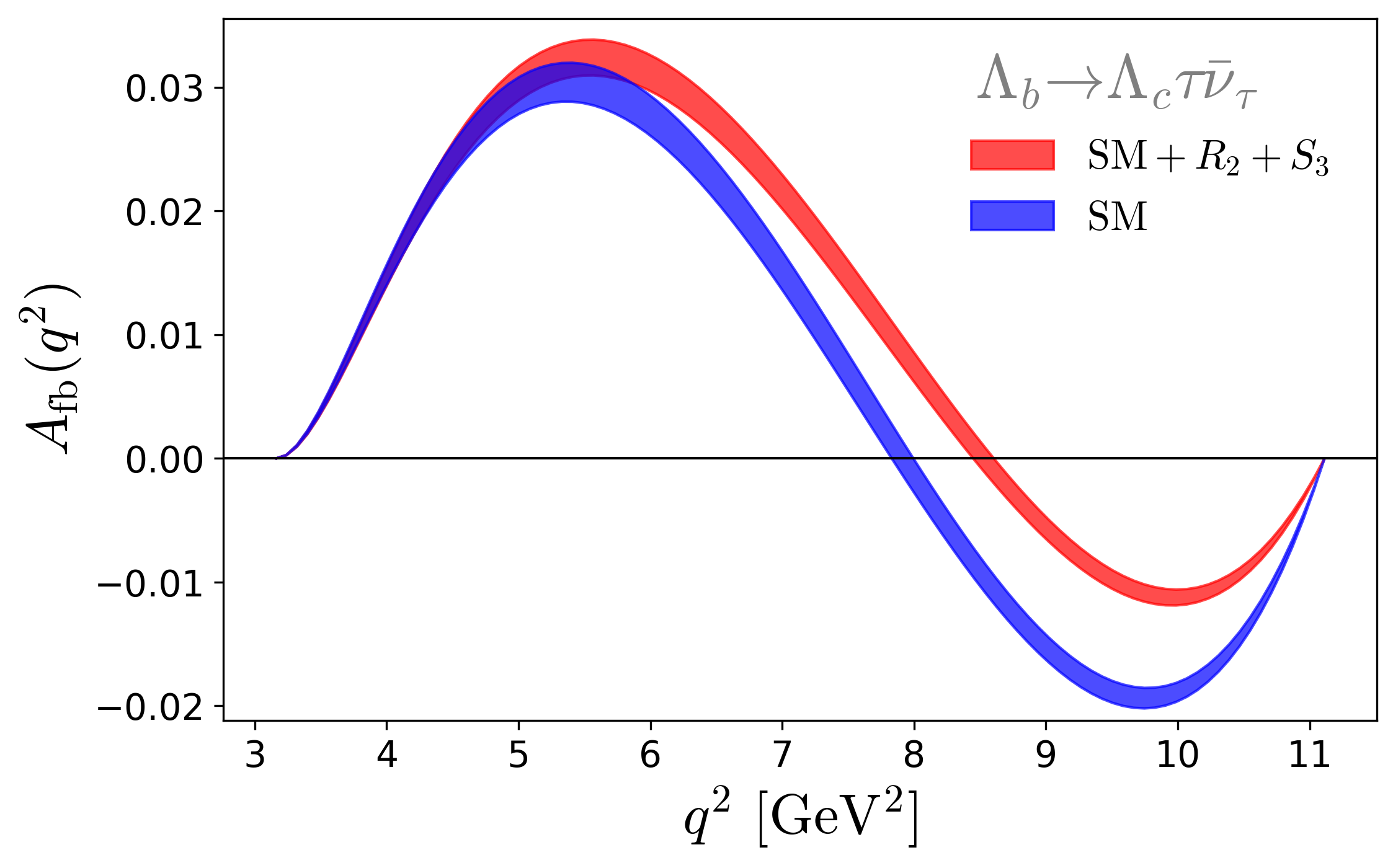}\\
  \vspace{5mm}
  \includegraphics[scale=0.14]{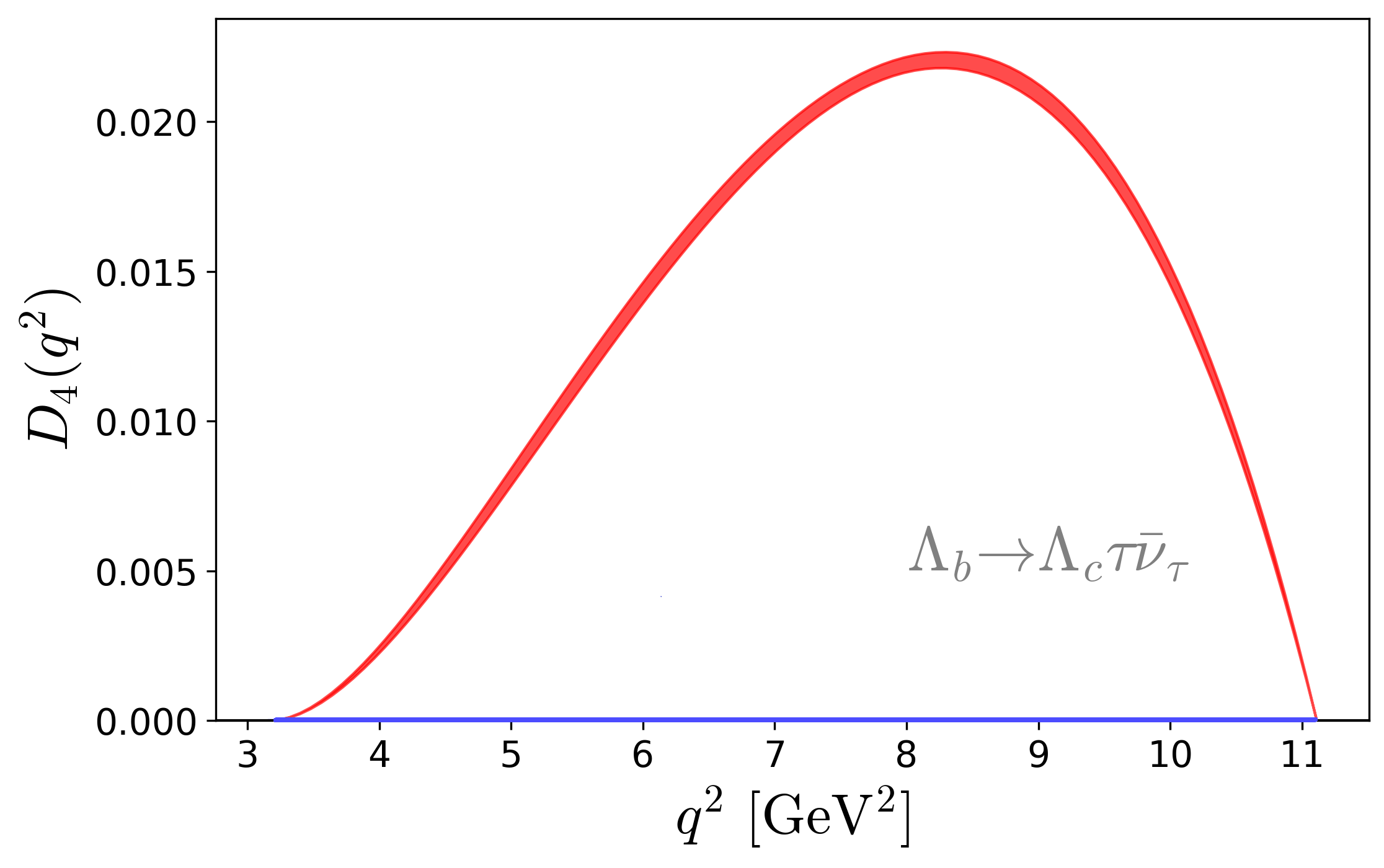}
  \caption{Two observables that can be extracted from the angular distribution of $\Lambda_b\to \Lambda_c(\to \Lambda \pi) \tau\nu$ and that have different $q^2$ shapes in the model presented here (depicted in red) from those in the SM (blue curves). Their values, after integrating in $q^2$ are in this model larger than in the SM. 
  \label{fig:fbd4}}
\end{figure}

The angular analysis of the exclusive $b\to c\tau \bar \nu$  modes can help identify several new observables, the measurement of which could help disentangle the situation and select among the currently viable scenarios. 
As an example we write the full angular distribution of the baryon decay as~\cite{Boer:2019zmp} 
\begin{equation}
\setlength{\jot}{10pt}
\begin{aligned}
&\frac{d^4\mathcal{B}}{dq^2d\cos\theta_\tau d \cos\theta  d\phi} = 8\pi\ \biggl[ \mathcal{A}_1 +\mathcal{A}_2 \cos\theta \biggr. \\ 
&\quad +\left(\mathcal{B}_1 +\mathcal{B}_2 \cos\theta \right)\cos\theta_\tau +\left(\mathcal{C}_1 +\mathcal{C}_2 \cos\theta \right)\cos^2\theta_\tau\\
&\quad +\left(\mathcal{D}_3 \sin\theta \cos\phi+\mathcal{D}_4 \sin\theta \sin\phi\right)\sin\theta_\tau\\
&\quad \biggl. +\left(\mathcal{E}_3 \sin\theta \cos\phi+\mathcal{E}_4 \sin\theta \sin\phi\right)\sin\theta_\tau\cos\theta_\tau \biggr],
\end{aligned}
\end{equation}
where the angles $\theta$ and $\theta_\tau$ are defined with respect to the direction of flight of $\Lambda_c$: $\theta$ being the angle of $\Lambda$  in the 
$\Lambda\pi$ rest frame, and $\theta_\tau$ is the angle of $\tau$ in the $\tau\bar \nu$-rest frame. $\phi$ is the angle between the $\tau \bar \nu$ and the $\Lambda\pi$ planes. 
In the above expression the $q^2$-dependent coefficient functions, $\mathcal{A}_{1,2}$, $\mathcal{B}_{1,2}$, $\mathcal{C}_{1,2}$, $\mathcal{D}_{3,4}$, $\mathcal{E}_{3,4}$, are given in terms of kinematical quantities and hadronic form factors~\cite{Boer:2019zmp}. Notice that all of the form factors relevant to any BSM discussion are already available, as they have all been  computed in lattice QCD away from the zero-recoil point~\cite{Detmold:2015aaa}. Forward-backward asymmetry is defined as
\bea
A_\mathrm{fb}(q^2)= \frac{1}{2}\, {  \mathcal{B}_1(q^2)\over   \Gamma(\Lambda_b\to \Lambda_c\tau\bar\nu)}\,,
\eea
where the full decay width is given by
\begin{align}
 \Gamma(\Lambda_b\to \Lambda_c\tau\bar\nu)= \int\displaylimits_{m_\tau^2}^{(m_{\Lambda_b}-m_{\Lambda_c})^2} dq^2\ \left[ \mathcal{A}_1(q^2) + \frac{1}{3} \mathcal{C}_1(q^2) \right] \,.
\end{align}
We find that for all of the available $g_{S_L}$ values discussed in the previous Section, the shape of $A_\mathrm{fb}(q^2)$ becomes different with respect to that found in the SM. In particular the point $q^2_0$, at which this asymmetry is zero, $A_\mathrm{fb}(q_0^2)=0$, is larger than the one found in the SM. Another quantity that one can use to monitor the viability of this model is 
\bea
D_4(q^2)=  {  \mathcal{D}_4(q^2)\over   \Gamma(\Lambda_b\to \Lambda_c\tau\bar\nu)}\,,
\eea
which is strictly zero in the SM and becomes non-zero only if the NP coupling can take a complex value, such as the case with our model, $\mathrm{Im}\left[ g_{S_L} \right] \neq 0$. 
In Fig.~\ref{fig:fbd4} we illustrate the change in shape of $A_\mathrm{fb}(q^2)$ and of $D_4(q^2)$ once $g_{S_L} =  8.1 \times g_T$ is switched to a plausible $g_{S_L} =0.5 i$.

We repeated the same exercise with $B\to D^\ast  \tau\nu$~\cite{angDstar} and found that the corresponding $A_\mathrm{fb}(q^2)$ changes only slightly. 
In order to support our observations by numerical values, we compute 
\begin{align}
\langle {O}  \rangle &= \int\displaylimits_{m_\tau^2}^{(M_{\Lambda_b}-M_{\Lambda_c})^2}\!\! O \ dq^2,
\end{align}
for $O \in \{ A_\mathrm{fb}, D_4\}$, and collect the results in Tab.~\ref{tab:1} where we also give the values for $q_0^2$ at which $A_\mathrm{fb}(q_0^2)=0$ and the results for the LFUV ratio
\begin{equation}
R_{\Lambda_c} = \dfrac{\mathcal{B}(\Lambda_b\to \Lambda_c \tau\bar{\nu})}{\mathcal{B}(\Lambda_b\to \Lambda_c \mu \bar{\nu})} .
\label{eq:RL_definition}
\end{equation}
From Tab.~\ref{tab:1} we see that $R_{\Lambda_c}$ follows the pattern and $R_{\Lambda_c} > R_{\Lambda_c}^\mathrm{SM}$. This can be tested with a more precise measurement of 
$R_{\Lambda_c}$. 
Furthermore, in this model we clearly observe that 
\begin{equation}
\langle {A_\mathrm{fb}}  \rangle > \langle {A_\mathrm{fb}}  \rangle^\mathrm{SM}, \qquad
\left| \langle {D_4}  \rangle \right|  > \left| \langle {D_4}  \rangle \right|^\mathrm{SM},
\end{equation}
which is in stark contrast with the models based on accommodating the $B$-anomalies by couplings to a $U_1$ vector LQ in which $\langle {A_\mathrm{fb}}  \rangle = \langle {A_\mathrm{fb}}  \rangle^\mathrm{SM}$, and $\langle {D_4}  \rangle = 0$. 
It is important to emphasize that these quantities can be used to discriminate among various scenarios proposed to explain $B$-anomalies.

\begin{table}[h]
\renewcommand{\arraystretch}{2.05}
\begin{tabular}{|c|c|c|}
\hline 
${\color{blue}g_{S_L}(m_b)}$ &     $0$ &          $-0.07\left({}^{+0.14}_{-0.14}\right) + 0.44\left({}^{+0.09}_{-0.12}\right)\, i$      \\\hline\hline
${\color{blue}R_{\Lambda_c}}$ &  $0.333(14)$ & $0.366(15)\left({}^{-0.002}_{+0.009}\right)\left({}^{+0.015}_{-0.014}\right)$ \\  \hline
${\color{blue}\langle {A_\mathrm{fb} }  \rangle }$ & $0.049(8)$& $0.085(7)\left({}^{+0.002}_{+0.004}\right)\left({}^{+0.014}_{-0.016}\right)$ \\ \hline
${\color{blue} q_0^2\,[\mathrm{GeV}^2] }$ & $7.97(7)$& $8.49(8)\left({}^{+0.00}_{+0.13}\right)\left({}^{+0.27}_{-0.25}\right)$  \\ \hline
${\color{blue} \langle {D_4 }  \rangle }$  &  $0$& $0.102(1)\left({}^{+0.001}_{-0.002}\right)\left({}^{+0.016}_{-0.025}\right)$ \\ \hline
\end{tabular}
\caption{ \sl \small Values of the observables relevant to $\Lambda_b\to \Lambda_c(\to \Lambda \pi) \tau\nu$, discussed in the text and computed in the SM ($g_{S_L}=0$) and for $g_{S_L}\neq 0$, as obtained from our scan, cf. Eq.~\eqref{eq:bfv}. Second and third uncertainties correspond to the variation of the central value with respect to the variation of the real and of the imaginary part of $g_{S_L}$, respectively.}
\label{tab:1} 
\end{table}

\section{Mass range for this scenario to remain valid}

So far in this paper we chose as a benchmark point the leptoquark masses $m_{R_2}=1.3$~TeV and $m_{S_3}=2$~TeV, consistent with the lower bounds deduced from the direct searches at the LHC, as discussed in Ref.~\cite{Angelescu:2018tyl}. From the low energy flavor physics observables we then obtained the constraints on the couplings of the model, and we pointed out that the very stringent constraints on the couplings can also be obtained from the analysis of the high-$p_T$ di-tau tails at the LHC. In order to monitor the range of masses preferred by this scenario we varied $m_{R_2}$ and $m_{S_3}$ and applied the same constraints on the couplings as before. We find that the model is highly sensitive to $m_{R_2}$, while it is only slightly sensitive to the variation of  $m_{S_3}$. 
\begin{figure}[!h]
  \centering
  \includegraphics[scale=0.57]{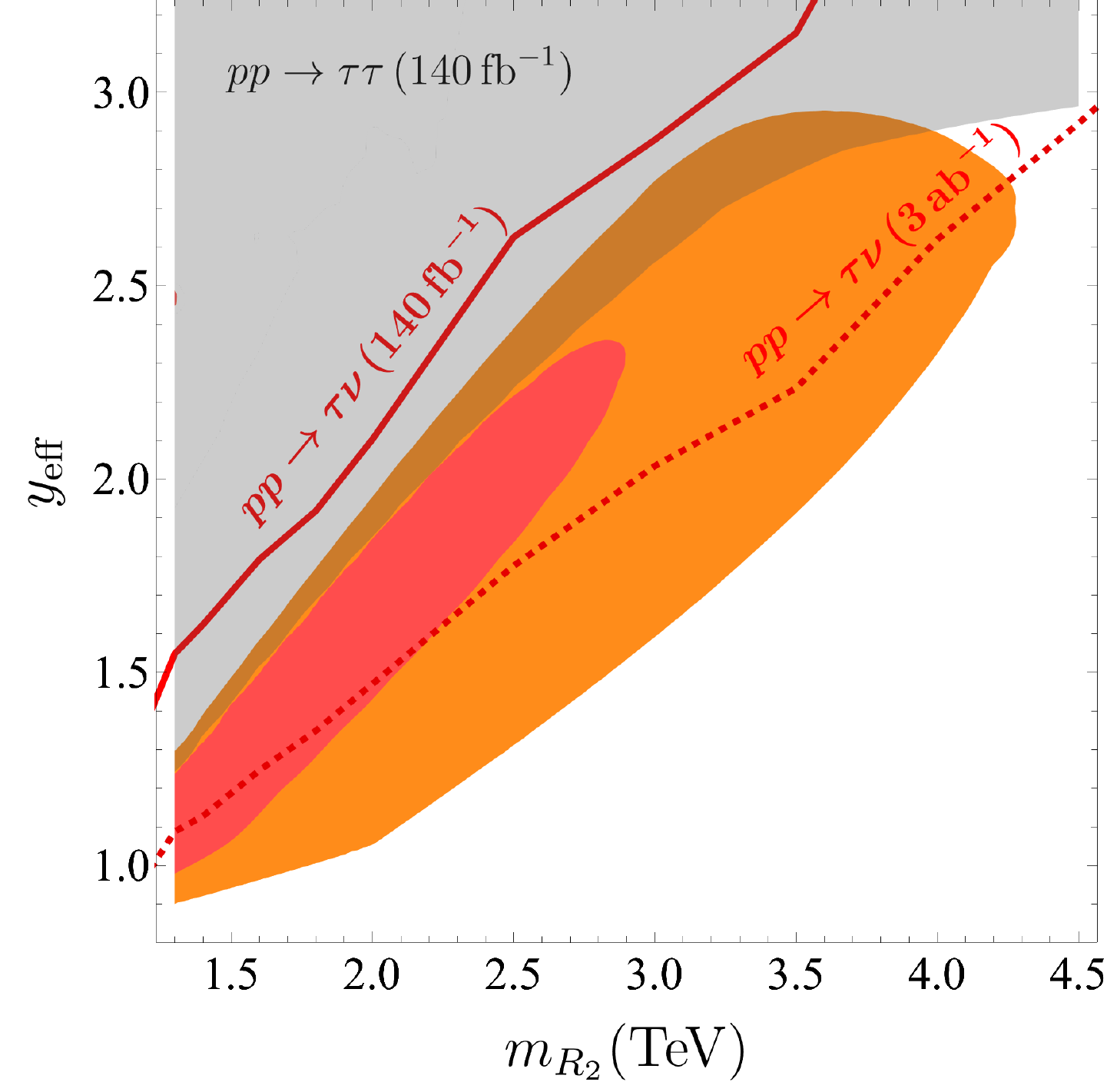}
  \caption{We plot the dependence of the effective coupling $y_\mathrm{eff} =\sqrt{\vert y_L^{c\tau} y_R^{b\tau\,\ast}\vert}$ on the variation of the leptoquark mass $m_{R_2}$. The orange regions are allowed by the low energy flavor physics constraints to $1\sigma$ and $2\sigma$. The gray area is excluded by the $2\sigma$ limits arising from the study of the high-$p_T$ tails of $pp\to \tau \tau$, as obtained from the currently available LHC data. We also plot the limit from the case of mono-tau in the final state.  
  \label{fig:mR2yeff}}
\end{figure}
The result is shown in Fig.~\ref{fig:mR2yeff} where we see that the current setup of Yukawa matrices, cf. Eqs.~(\ref{eq:yL-yR},\ref{eq:yL-S3}), remain consistent with the constraints to $2\sigma$ if $m_{R_2}\lesssim 4.3$~TeV. In other words, if the flavor constraints remain unchanged, this scenario can be tested at the LHC. 
It is also interesting to note from Fig.~\ref{fig:mR2yeff} that the effective coupling $y_\mathrm{eff} =\sqrt{\vert y_L^{c\tau} y_R^{b\tau\,\ast}\vert}$ always remains well below the perturbativity limit, $y_\mathrm{eff}\leq \sqrt{4\pi}$.

\section{Conclusion}
We update the parameter space of the model in which the SM is extended by $\mathcal{O}(1\,\mathrm{TeV})$ two scalar LQ's, $R_2$ and $S_3$, and show that this model is still a plausible framework to accommodate the $B$-anomalies while remaining consistent both with a number of experimental constraints arising from the low energy observables,  as well as with those deduced from the LHC measurements relevant to the high-$p_T$ tails of $pp\to \tau \tau$ and $pp\to \tau \nu$. A peculiarity of this $R_2$-$S_3$ scenario is that there is a complex coupling. We find that the size of the corresponding imaginary part of the model parameter $y_R^{b\tau}\propto g_{S_L}$ results in: (i) a value of the electric dipole moment of the neutron consistent with the experimental bound, (ii) too small a contribution to $\Delta a_\mathrm{CP}$, difference of the CP-asymmetries between $D^0\to K^+K^-$ and $D^0\to \pi^+\pi^-$, (iii) a significant change in the observables that can be deduced from the angular distribution of $B\to D^\ast (\to D\pi) \tau\nu$ and $\Lambda_b\to \Lambda_c(\to \Lambda \pi) \tau\nu$ and which are zero in the SM and in scenarios in which the NP couplings are real. We also find that the forward-backward asymmetry in the case of $\Lambda_b\to \Lambda_c \tau\nu$ becomes significantly different from its SM value. Like in the other models built to accommodate $B$-anomalies and involving LQ's, we establish the 
upper and lower bounds to the exclusive LFV decay modes based on $b\to s \mu^\pm \tau^\mp$. We also checked that the model gives a negligible contribution to $\mathcal{B}(K\to\pi \nu\nu)$, but it significantly enhances $\mathcal{B}(B\to K^{(\ast )} \nu\nu)$, cf. \eqref{eq:nunu}, which will soon be experimentally scrutinized at Belle-II. We also find a clear correlation between $\mathcal{B}(B\to K^{(\ast )} \nu\nu)$ and the LFV decays such as $\mathcal{B}(B\to K \mu\tau)$. Importantly, the model remains consistent with the current experimental upper bound on $\mathcal{B}(\tau \to \mu \gamma)$.

\begin{acknowledgments}
S.F. and N.K.\ acknowledge support of the Slovenian Research Agency under the core funding grant P1-0035. D.A.F has received funding from the European Research Council (ERC) under the European Union's Horizon 2020 research and innovation programme under grant agreement 833280 (FLAY), and by the Swiss National Science Foundation (SNF) under contract 200021-175940. This work has also been supported in part by Croatian Science Foundation under the project 7118 and the European Union's Horizon 2020 research and innovation programme under the Marie Sklodowska-Curie grant agreement N$^\circ$~660881-Hidden.
\end{acknowledgments}

\end{document}